\documentclass[epj]{svjour}
\usepackage{graphics}
\usepackage{graphicx}

\begin{document}
\title{NA60 results on thermal dimuons}

\subtitle{NA60 Collaboration}
\author{
R.~Arnaldi\inst{11} \and
K.~Banicz\inst{4,6} \and
K.~Borer\inst{1} \and
J.~Castor\inst{5} \and 
B.~Chaurand\inst{9} \and 
W.~Chen\inst{2} \and
C.~Cical\`o\inst{3} \and  
A.~Colla\inst{11} \and 
P.~Cortese\inst{11} \and
S.~Damjanovic\inst{4} 
\thanks{\emph{Corresponding author:} sanja.damjanovic@cern.ch} \and
A.~David\inst{4,7} \and 
A.~de~Falco\inst{3} \and 
A.~Devaux\inst{5} \and 
L.~Ducroux\inst{8} \and 
H.~En'yo\inst{10} \and
J.~Fargeix\inst{5} \and
A.~Ferretti\inst{11} \and 
M.~Floris\inst{3} \and 
A.~F\"orster\inst{4} \and
P.~Force\inst{5} \and
N.~Guettet\inst{4,5} \and
A.~Guichard\inst{8} \and 
H.~Gulkanian\inst{12} \and 
J.M.~Heuser\inst{10} \and
M.~Keil\inst{4,7} \and 
L.~Kluberg\inst{9} \and 
Z.~Li\inst{2} \and
C.~Louren\c{c}o\inst{4} \and
J.~Lozano\inst{7} \and 
F.~Manso\inst{5} \and 
P.~Martins\inst{4,7} \and  
A.~Masoni\inst{3} \and
A.~Neves\inst{7} \and 
H.~Ohnishi\inst{10} \and 
C.~Oppedisano\inst{11} \and
P.~Parracho\inst{4,7} \and 
P.~Pillot\inst{8} \and 
T.~Poghosyan\inst{12} \and
G.~Puddu\inst{3} \and 
E.~Radermacher\inst{4} \and
P.~Ramalhete\inst{4,7} \and 
P.~Rosinsky\inst{4} \and 
E.~Scomparin\inst{11} \and
J.~Seixas\inst{7} \and 
S.~Serci\inst{3} \and 
R.~Shahoyan\inst{4,7} \and 
P.~Sonderegger\inst{7} \and
H.J.~Specht\inst{6} \and 
R.~Tieulent\inst{8} \and 
G.~Usai\inst{3} \and 
R.~Veenhof\inst{7} \and
H.K.~W\"ohri\inst{3,7}.
}

\institute{
Laboratory for High Energy Physics, Bern, Switzerland. \and  
BNL, Upton, New York, USA. \and                              
Universit\`a di Cagliari and INFN, Cagliari, Italy. \and     
CERN, Geneva, Switzerland. \and                              
LPC, Universit\'e Blaise Pascal and CNRS-IN2P3, Clermont-Ferrand,
  France. \and                                               
Physikalisches Institut der Universit\"{a}t Heidelberg,
  Germany. \and                                              
IST-CFTP, Lisbon, Portugal. \and                             
IPN-Lyon, Univ.\ Claude Bernard Lyon-I and CNRS-IN2P3,
  Lyon, France. \and                                         
LLR, Ecole Polytechnique and CNRS-IN2P3, Palaiseau,
  France. \and                                               
RIKEN, Wako, Saitama, Japan. \and                            
Universit\`a di Torino and INFN, Italy. \and                 
YerPhI, Yerevan, Armenia.                                    
}

%
%
%
\date{Received: date / Revised version: date}
%

\abstract{ The NA60 experiment at the CERN SPS has measured muon pairs
with unprecedented precision in 158A GeV In-In collisions. A strong
excess of pairs above the known sources is observed in the whole mass
region 0.2$<$$M$$<$2.6 GeV. The mass spectrum for $M$$<$1 GeV is
consistent with a dominant contribution from $\pi^{+}\pi^{-}
\rightarrow \rho \rightarrow \mu^{+}\mu^{-}$ annihilation.  The
associated $\rho$ spectral function shows a strong broadening, but
essentially no shift in mass. For $M$$>$1 GeV, the excess is found to
be prompt, not due to enhanced charm production, with pronounced
differences to Drell-Yan pairs. The slope parameter $T_\mathrm{eff}$
associated with the transverse momentum spectra rises with mass up to
the $\rho$, followed by a sudden decline above. The rise for $M$$<$1
GeV is consistent with radial flow of a hadronic emission source. The
seeming absence of significant flow for $M$$>$1 GeV and its relation
to parton-hadron duality is discussed in detail, suggesting a
dominantly partonic emission source in this region. A comparison of
the data to the present status of theoretical modeling is also
contained. The accumulated empirical evidence, including also a
Planck-like shape of the mass spectra at low $p_{T}$ and the lack of
polarization, is consistent with a global interpretation of the excess
dimuons as thermal radiation. We conclude with first results on
$\omega$ in-medium effects.
} 
\authorrunning{NA60 Collaboration}
\titlerunning{NA60 results on thermal dimuons}
\maketitle
\section{Introduction}
\label{intro}
Dileptons are particularly attractive to study the hot and dense QCD
matter formed in high-energy nuclear collisions. In contrast to
hadrons, they directly probe the entire space-time evolution of the
expanding system, escaping freely without final-state interactions. At
low masses $M$$<$1 GeV (LMR), thermal dilepton production is mediated
by the broad vector meson $\rho$ (770) in the hadronic phase. Due to
its strong coupling to the $\pi\pi$ channel and the short life time of
only 1.3 fm/c, ``in-medium'' modifications of its mass and width close
to the QCD phase boundary have since long been considered as the prime
signature for {\it chiral symmetry
restoration}~\cite{Pisarski:mq,Rapp:1995zy,Brown:kk}. At intermediate
masses $M$$>$1 GeV (IMR), it has been controversial up to today
whether thermal dileptons are dominantly produced in the earlier
partonic or in the hadronic phase, based here on hadronic processes
other than $\pi\pi$ annihilation. Originally, thermal emission from
the early phase was considered as a prime probe of {\it
deconfinement}~\cite{McLerran:1984ay,Kajantie:1986}.

Experimentally, it took more than a decade to master the challenges of
very rare signals and enormous combinatorial backgrounds. The first
clear signs of an excess of dileptons above the known decay sources at
SPS energies were obtained by
CERES~\cite{Agakichiev:1995xb,CERES:2008} for $M$$<$1 GeV,
NA38/NA50~\cite{Abreu:2002rm} for $M$$>$1 GeV and by
HELIOS-3~\cite{Masera:1995ck} for both mass regions
(see~\cite{Specht:2007ez} for a short recent review including the
preceding $pp$ era and the theoretical milestones). The sole existence
of an excess gave a strong boost to theory, with hundreds of
publications. In the LMR region, $\pi\pi$ annihilation with
regeneration and strong in-medium modifications of the intermediate
$\rho$ during the fireball expansion emerged as the dominant
source. However, the data quality in terms of statistics and mass
resolution remained largely insufficient for a precise assessment for
the in-medium spectral properties of the $\rho$. In the IMR region,
thermal sources or enhanced charm production could account for the
excess equally well, but that ambiguity could not be resolved, nor
could the nature of the thermal sources be clarified.

\begin{figure}[t]
\begin{center}
\includegraphics*[width=0.43\textwidth]{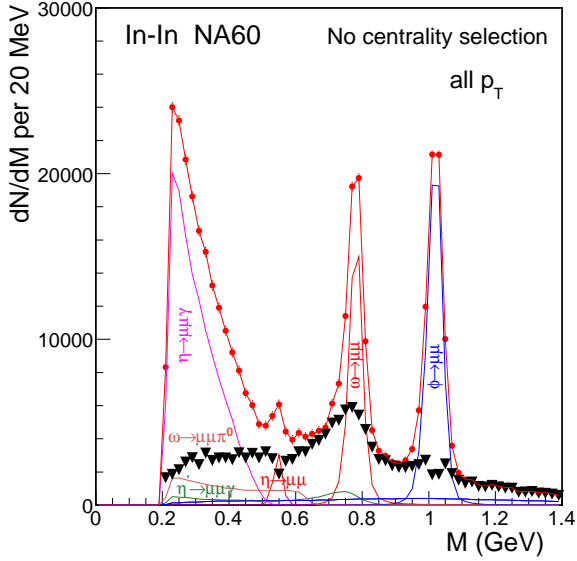}
\caption{Background-subtracted mass spectrum before (dots) and after
subtraction of the known decay sources (triangles).}
   \label{fig1}
\end{center}
\end{figure}

A big step forward in technology, leading to completely new standards
of the data quality in this field, has recently been achieved by NA60,
a third-generation experiment built specifically to follow up the open
issues addressed above~\cite{NA60:2008IMR}. Initial results on mass
and transverse momentum spectra of the excess dimuons have already
been published~\cite{NA60:2008IMR,Arnaldi:2006jq,Arnaldi:2007ru},
supplemented by recent results on acceptance-corrected mass spectra
and polarization~\cite{Damjanovic:2008ta}. This paper takes a broader
view on the results, discussing in some detail the observed
$M$-$p_{T}$ correlations, the connection to hadron-parton duality, and
the present status of theoretical modeling. Certain aspects of the
$p_{T}$ spectra, of centrality dependencies and evidence for $\omega$
in-medium effects are reported here for the first time.

\section{Mass spectra and the $\rho$ spectral function}
\label{sec:1}

Fig.~\ref{fig1} shows the centrality-integrated net dimuon mass
spectrum for 158A GeV In-In collisions in the LMR region. The narrow
vector mesons $\omega$ and $\phi$ are completely resolved; the mass
resolution at the $\omega$ is 20 MeV. The peripheral data can be
completely described by the electromagnetic decays of neutral
mesons~\cite{Arnaldi:2006jq,Damjanovic:2006bd}. This is not true for
the more central data as plotted in Fig.~\ref{fig1}, due to the
existence of a strong excess of pairs. The high data quality of NA60
allows to isolate this excess with {\it a priori unknown
characteristics} without any fits: the cocktail of the decay sources
is subtracted from the total data using {\it local} criteria, which
are solely based on the mass distribution itself. The $\rho$ is not
subtracted. The excess resulting from this difference formation is
illustrated in the same figure
(see~\cite{Arnaldi:2006jq,Arnaldi:2007ru,Damjanovic:2006bd} for
details and error discussion). The subtracted data for the $\eta$,
$\omega$ and $\phi$ themselves are subject to the same further steps
as the excess data and are used later for comparison.

\begin{figure}[t]
\begin{center}
\includegraphics*[width=0.43\textwidth]{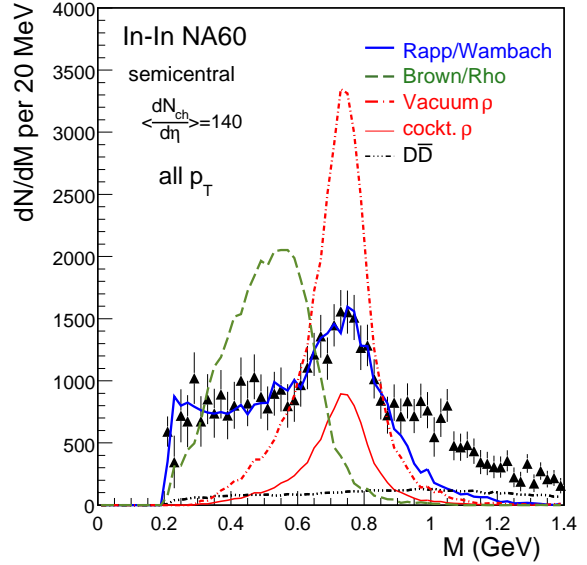}
\caption{Excess dimuons compared to theoretical
predictions~\cite{Rapp:2005pr}, renormalized to the data in the mass
interval $M$$<$0.9 GeV. No acceptance correction applied.}
   \label{fig2}
\end{center}
\end{figure}

\begin{figure*}[]
\centering
\includegraphics*[width=0.46\textwidth]{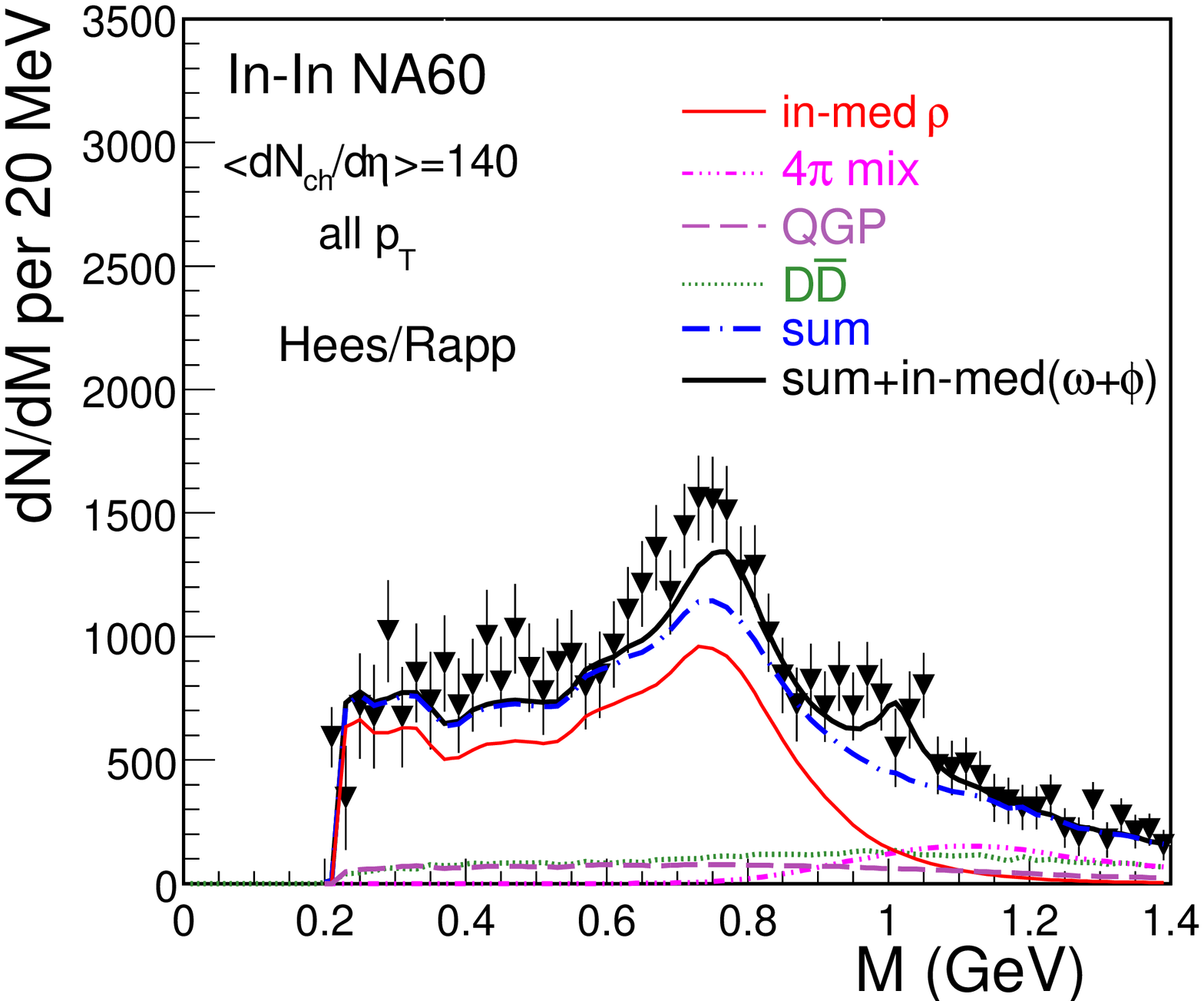}
\includegraphics*[width=0.46\textwidth]{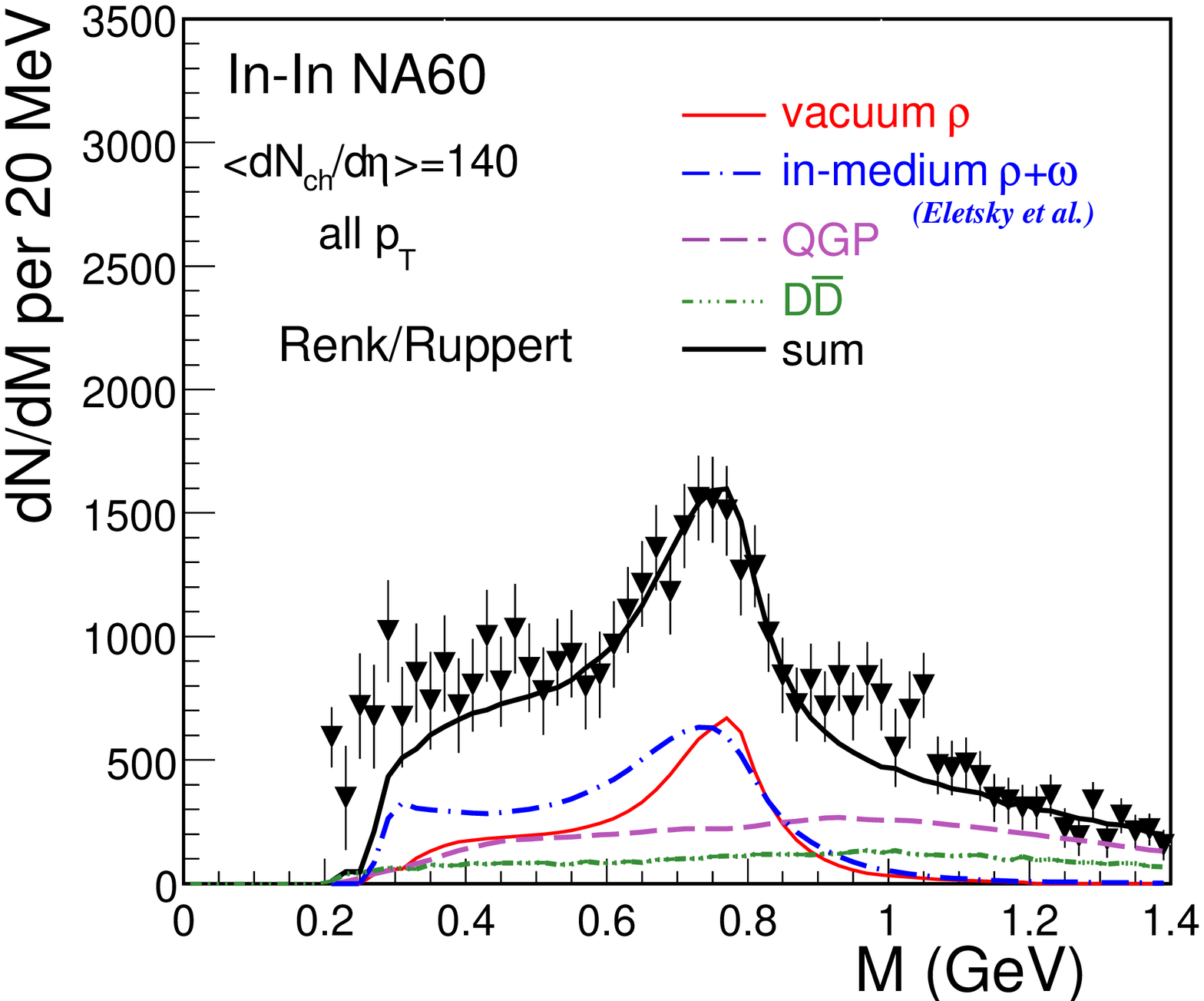}
\caption{Excess dimuons for semicentral collisions compared to the
theoretical model results by Hees/Rapp~\cite{vanHees:2006ng} (left)
and Renk/Ruppert et al.~\cite{Ruppert:2007cr} (right). No acceptance
correction applied.}
\label{fig3}
\end{figure*}

The common features of the excess mass spectra can be recognized in
Fig.~\ref{fig2}. A peaked structure is always seen, residing on a
broad continuum with a yield strongly increasing with centrality (see
Fig.~\ref{fig9} below), but remaining essentially centered around the
nominal $\rho$ pole~\cite{Damjanovic:2006bd}. Without any acceptance
correction and $p_{T}$ selection, the data can directly be interpreted
as the {\it space-time averaged spectral function} of the $\rho$, due
to a fortuitous cancellation of the mass and $p_{T}$ dependence of the
acceptance filtering by the phase space factors associated with
thermal dilepton emission~\cite{Damjanovic:2006bd}. The two main
theoretical scenarios for the in-medium spectral properties of the
$\rho$, broadening~\cite{Rapp:1995zy} and dropping
mass~\cite{Brown:kk}, are shown for comparison, both evaluated for the
same fireball evolution~\cite{Rapp:2005pr}. Since agreement between
modeling and data would imply agreement both in shape and yield, the
model results are normalized to the data in the mass interval $M$$<$
0.9 GeV, just to be independent of the uncertainties of the fireball
evolution. The unmodified $\rho$, also shown in Fig.~\ref{fig2}
(vacuum $\rho$), is clearly ruled out. The broadening scenario indeed
gets close, while the dropping mass scenario in the version which
described the CERES data reasonably
well~\cite{Rapp:1995zy,Brown:kk,Agakichiev:1995xb} completely fails
for the much more precise NA60 data. A strong reduction of in-medium
VMD as proposed by the vector manifestation of chiral
symmetry~\cite{Harada:2007zw} would make hadron spectral functions in
hot and dense matter altogether unobservable, but central aspects of
this scenario are totally unclear, and quantitative predictions which
could be confronted with data have not become available up to today.

A comparison of the same excess mass spectrum to two more recent
theoretical developments, covering now both the LMR and the initial
part of the IMR region, is contained in Fig.~\ref{fig3}. In contrast
to Fig.~\ref{fig2}, the theoretical results are not renormalized here,
but shown on an absolute scale. In the $\rho$-like region with
$\pi^{+}\pi^{-} \rightarrow \rho \rightarrow \mu^{+}\mu^{-}$ as the
dominant source, Hees/Rapp~\cite{vanHees:2006ng} use the original
many-body scenario with a $\rho$ spectral function strongly broadened by
baryonic interactions~\cite{Rapp:1995zy}, while Renk/Ruppert's
results~\cite{Ruppert:2007cr} are based on the spectral function of
Eletsky et al.~\cite{Eletsky:2001bb} where the broadening effects from
baryons are somewhat weaker; that difference is directly visible in
the low-mass tails of the theoretical mass spectra. The overall
agreement between the data and the two theoretical scenarios is quite
satisfactory in this region, also in absolute terms.

\section{Mass spectra and parton-hadron duality}
\label{sec:2}

Moving up into the IMR region $M$$>$1 GeV, 2$\pi$ processes become
negligible, and other hadronic processes like 4$\pi$ (including
vector-axialvector mixing in case of ~\cite{vanHees:2006ng}) and
partonic processes like quark-antiquark annihilation $q \bar{q}
\rightarrow \mu^{+}\mu^{-}$ take over. The two theoretical scenarios
in Fig.~\ref{fig3} also describe this part. However, there is a very
interesting and instructive difference between them. While the total
yield of the data for $M$$>$1 GeV is described about equally well, the
fraction of partonic processes relative to the total is small
in~\cite{vanHees:2006ng} where a first-order phase transition is used,
and dominant in~\cite{Ruppert:2007cr} which uses a cross-over phase
transition. This feature is often referred to as ``parton-hadron
duality'' and formed the basis of the successful description of the
NA50 dimuon enhancement in the IMR region in terms of thermal
radiation~\cite{Rapp:1999zw}. Here, the individual sources were not
even specified.

Caution should, however, be expressed as to the use of the term
``duality'' in this context. Parton-hadron duality is a statement on
dilepton emission {\it rates}, dating back to the time-reversed
process of hadron production in $e^{+}e^{-}$ collisions. It implies
that the emission rates using either partonic (pQCD) or hadronic
degrees of freedom merge together, i.e. become ``dual'', if the system
approaches deconfinement and chiral restoration. The validity of
duality down to masses of 1 GeV, mainly due to vector-axialvector
mixing, was first shown by Li and Gale~\cite{CGale:1998ej} (see
also~\cite{Rapp:1995zy}). However, experiments measure {\it yields},
i.e. rates integrated over space-time. Duality in the yields is not
obvious and becomes questionable, if the space-time trajectories are
{\it different} for genuine partonic and hadronic processes. Such a
difference automatically appears through the elementary assumption
that partonic processes only act ``early'', i.e. from $T_{i}$ until
$T_{c}$, while hadronic processes (like $n \pi$) only act ``late'',
i.e. from $T_{c}$ to thermal freeze-out $T_{f}$. If theoretical
scenarios are {\it different} in their trajectories (both as to
partonic and to hadronic processes), the integrated total yields will,
in general, be different. Since the scenarios of~\cite{vanHees:2006ng}
and~\cite{Ruppert:2007cr} in Fig.~\ref{fig3} are indeed very different
(see above), the seemingly equivalent description of the data cannot
be traced to duality, but must be due to internal parameter choices.

Explicit insight beyond duality, be it real or fortuitous, can be
obtained experimentally in the following way (see~\cite{Kajantie:1986}
and in particular~\cite{Renk:2008prc}). In contrast to real photons,
virtual photons decaying into lepton pairs are characterized by two
variables, mass $M$ and transverse momentum $p_{T}$. Historically, the
interest has largely focused on mass because of its rich and often
structured information content, including now the $\rho$ spectral
function discussed above. Transverse momentum, on the other hand,
contains not only contributions from the spectral function(s), but
encodes the key properties of the expanding fireball, temperature and,
in particular, transverse (radial) flow. The latter causes a blue
shift of $p_{T}$, analogous to the case of hadrons. In contrast to
hadrons, however, which always receive the full asymptotic flow
reached at the moment of decoupling from the flowing medium, lepton
pairs are continuously emitted during the evolution, reflecting a
space-time folding over the temperature-flow history in their final
$p_{T}$ spectra. Since flow builds up monotonically during the
evolution, being small in the early partonic phase (at SPS energies,
due to the ``soft point'' in the equation-of-state), and increasingly
larger in the late hadronic phase, the final $p_{T}$ spectra keep
memory on the {\it time ordering} of the different dilepton sources,
mirrored in a characteristic {\it mass dependence of the $p_{T}$
spectra}. We shall come back to this point below.

\section{Acceptance-corrected mass and $p_{T}$ spectra}
\label{sec:3}

Quantitative insight into the physical meaning of the excess dileptons
requires a full correction of the data for geometrical acceptance and
pair efficiencies of the NA60 apparatus, including the effects of the
trigger system. Results from Monte Carlo simulations of the acceptance
are contained in~\cite{Arnaldi:2007ru,Damjanovic:2007qm}, showing
significant variations and in particular a strong decrease at low mass
and low $p_{T}$. In principle, the correction requires a 4-dimensional
grid in the space of $M$-$p_{T}$-$y$-$cos{\theta_{CS}}$ (where
$\theta_{CS}$ is the polar angle of the muons in the Collins Soper
frame). To avoid large statistical errors in low-acceptance bins, it
is performed instead in 2-dimensional $M$-$p_{T}$ space, using the
measured $y$ and $cos{\theta}$ distributions as an input. The latter
are, in turn, obtained with acceptance corrections determined in an
iterative way from MC simulations matched to the data in $M$ and
$p_{T}$. The $y$-distribution is found to have the same rapidity width
as $dN_{ch}/d\eta$, $\sigma_{y}\sim$ 1.5~\cite{Damjanovic:2007qm}. The
$cos{\theta_{CS}}$ distributions for two mass windows of the excess
and the $\omega$ are contained in~\cite{Damjanovic:2008ta}. Within
errors, they are found to be uniform, implying the polarization of the
excess dimuons to be zero, in contrast to Drell-Yan and consistent
with the expectations for thermal radiation from a randomized system.

\begin{figure*}[htbp]
\begin{center}
\includegraphics*[width=0.29\textwidth]{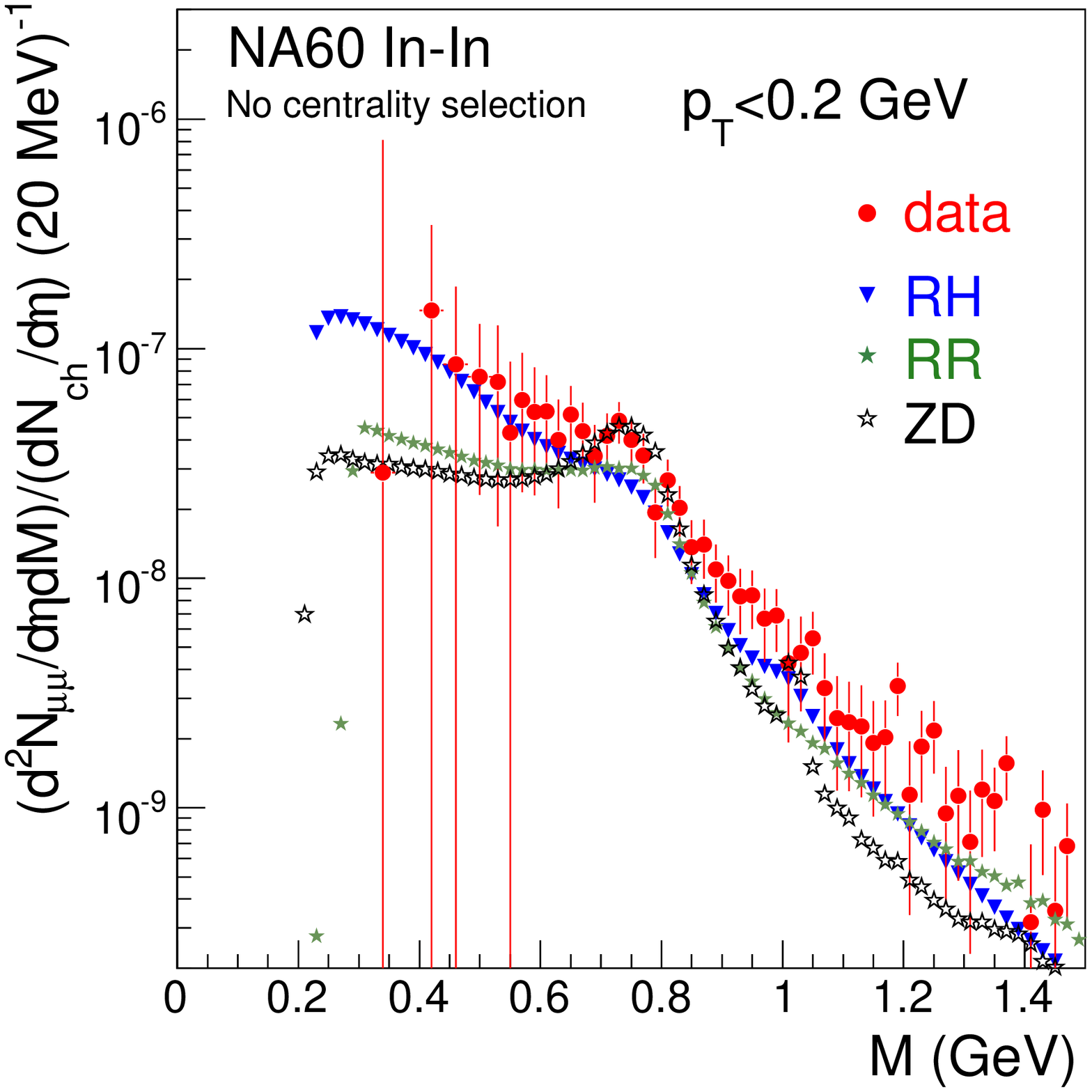}
\includegraphics*[width=0.29\textwidth]{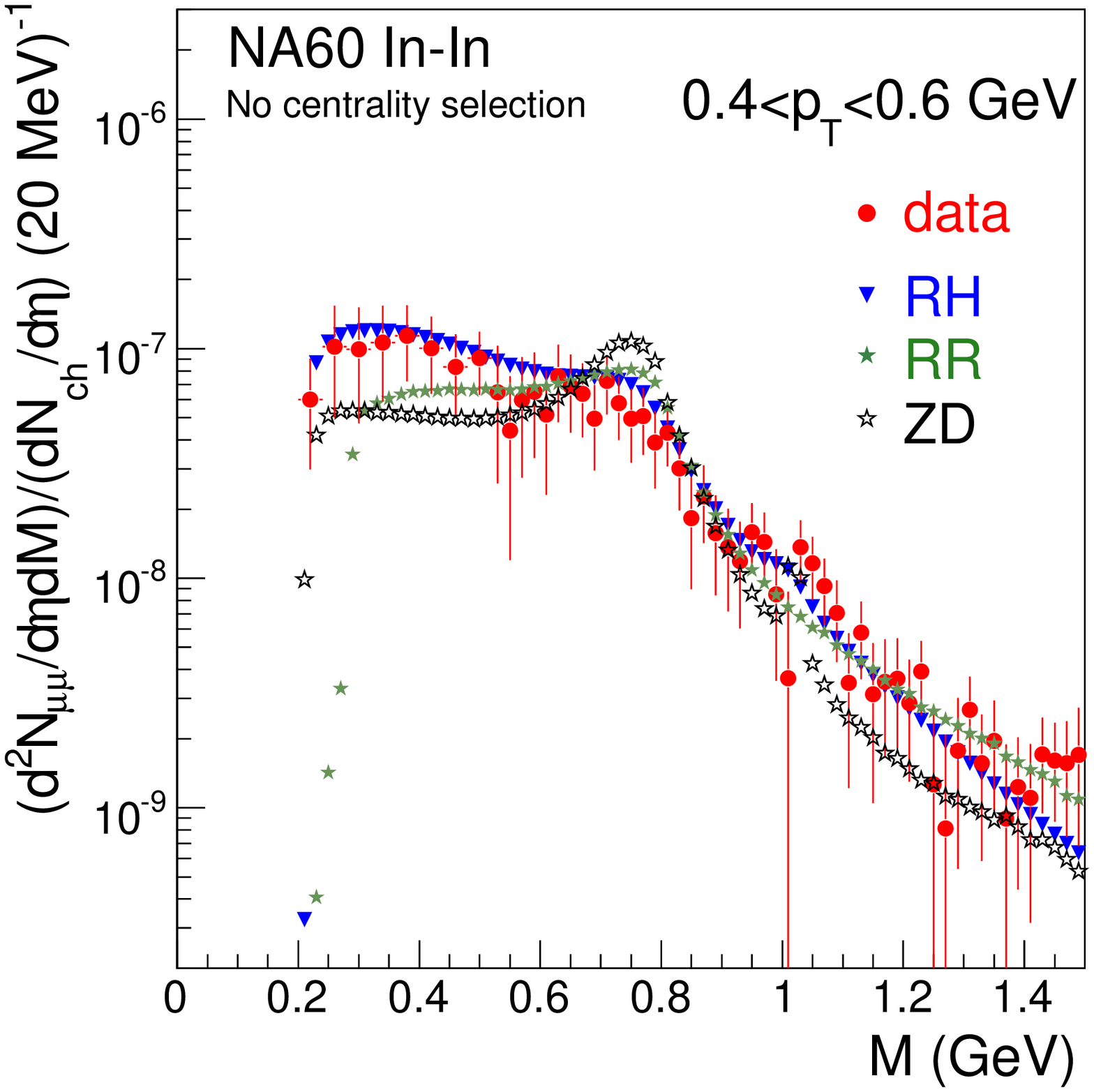}
\includegraphics*[width=0.29\textwidth]{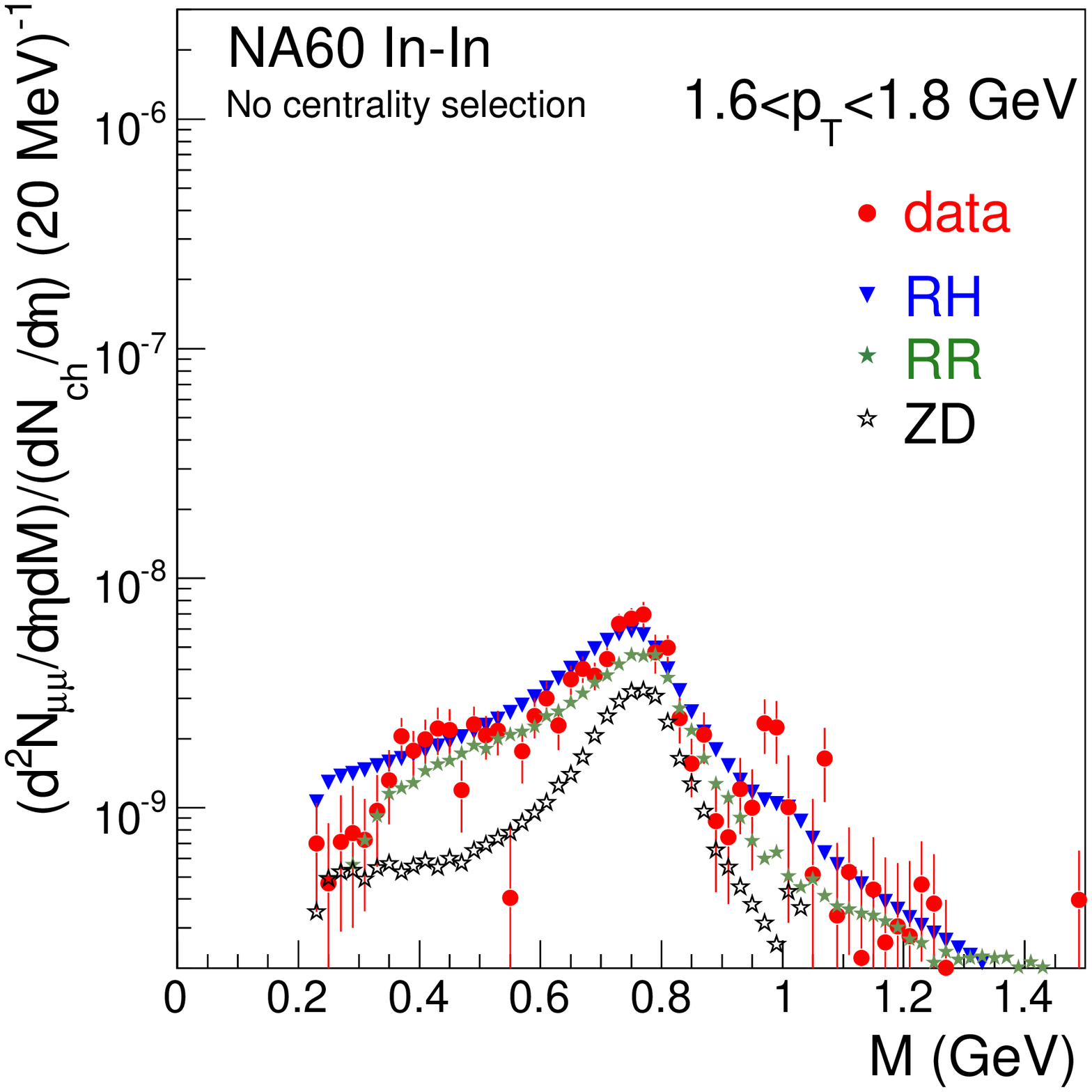}
\vglue -1.5mm
\includegraphics*[width=0.29\textwidth]{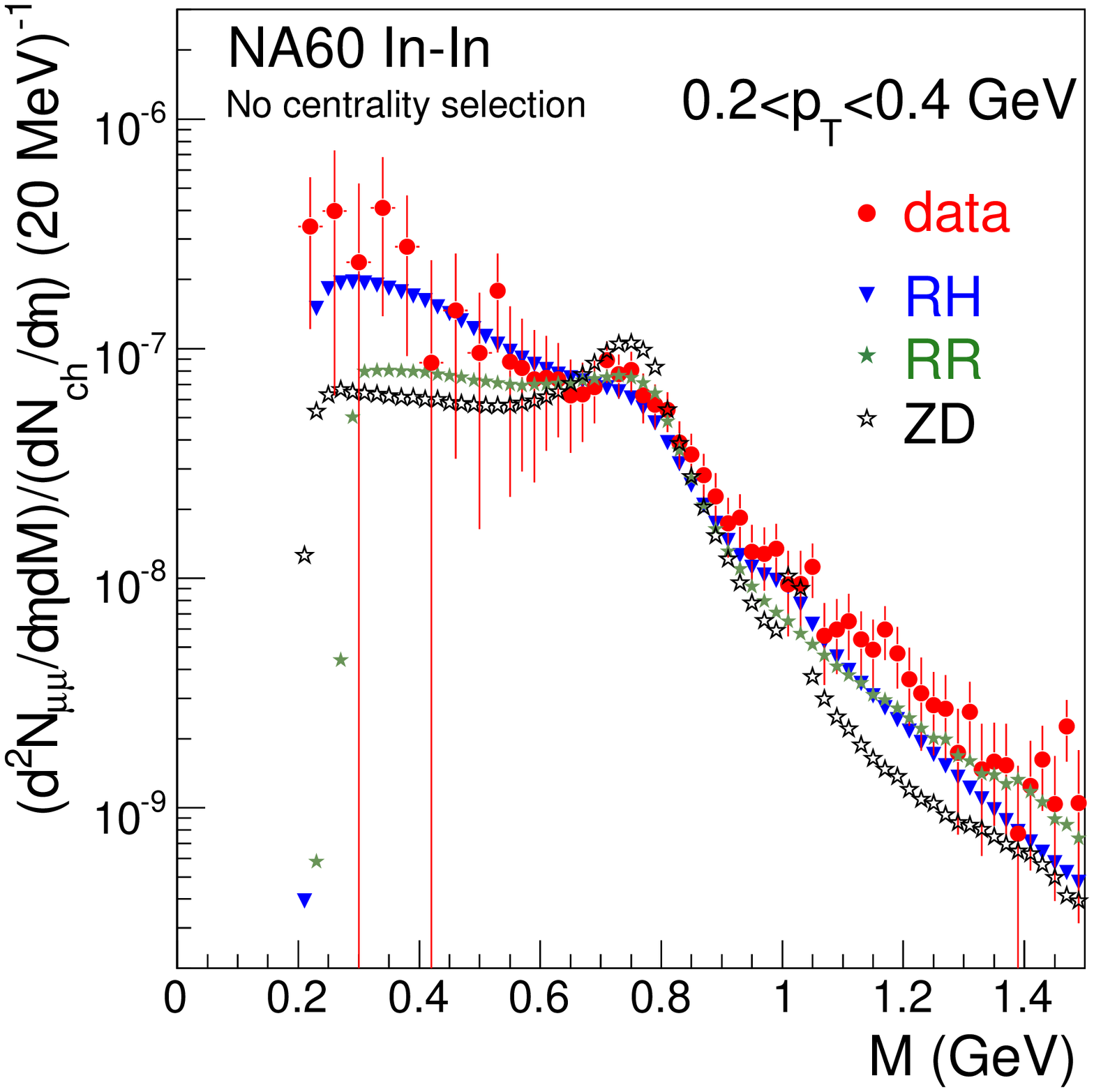}
\includegraphics*[width=0.29\textwidth]{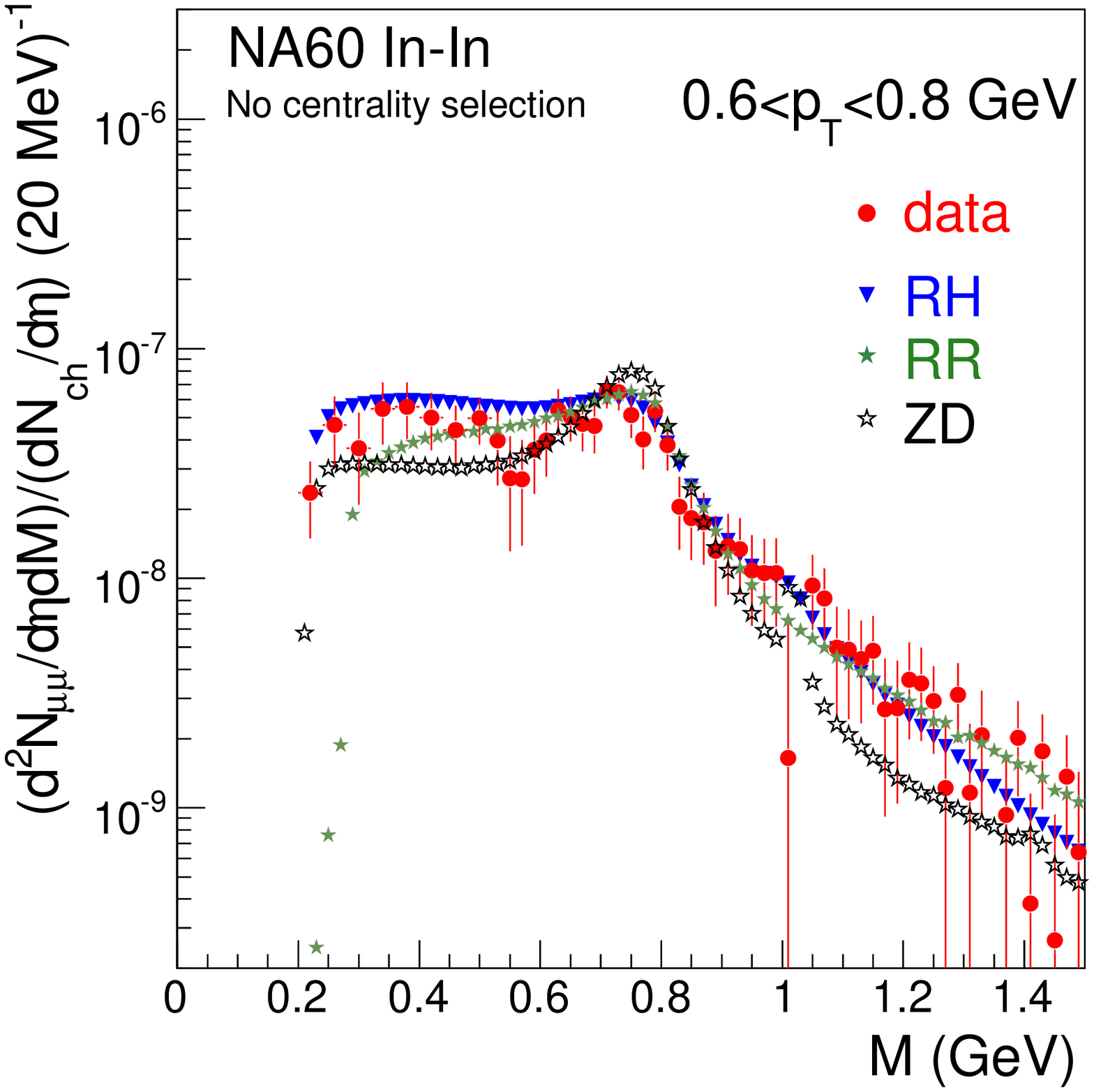}
\includegraphics*[width=0.29\textwidth]{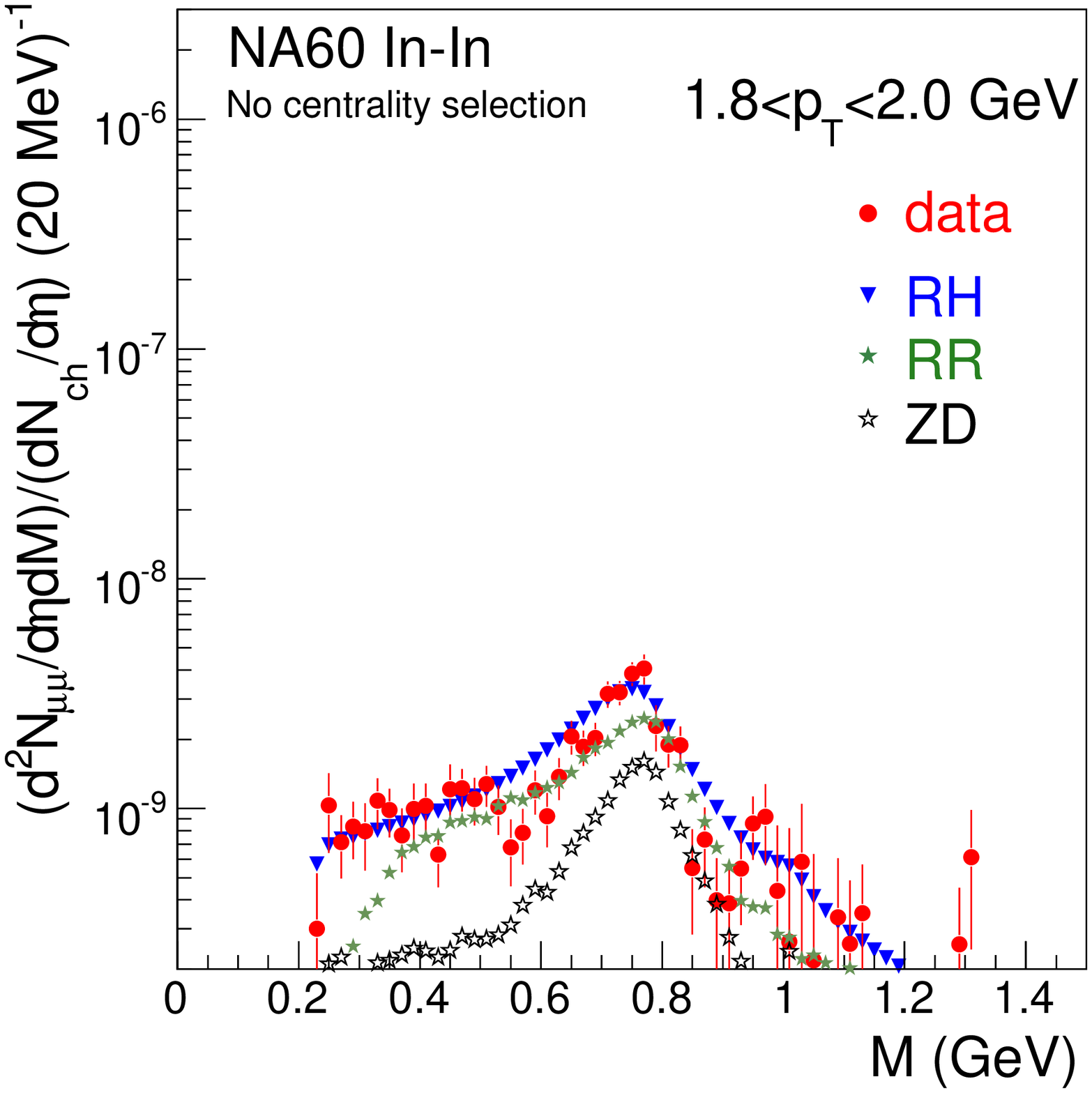}
\caption{Acceptance-corrected mass spectra of the excess dimuons in
selected slices of $p_{T}$. Absolute normalization as in
Fig.~\ref{fig4}. The theoretical scenarios are labeled according to
the authors HR~\cite{RH:2008lp}, RR~\cite{Renk:2008prc}, and
ZD~\cite{Dusling:2007rh}. In case of~\cite{RH:2008lp}, the EoS-B$^{+}$
option is used, leading to a partonic fraction of about 65\% in the
IMR (different from the left part of Fig.~\ref{fig3}).}
   \label{fig4}
\end{center}
\end{figure*}

The outcome for the two major variables $M$ and $p_{T}$ is first
discussed separately for the LMR and the IMR regions. Fig.~\ref{fig4}
shows a set of mass spectra for some selected slices in $p_{T}$ to
illustrate the evolution from low to high $p_{T}$ (a
$p_{T}$-integrated mass spectrum over the whole mass region is
contained in~\cite{NA60:2008IMR}). The spectra are normalized to
$dN_{ch}/d\eta$ in absolute terms, using the same procedure as
described in detail for the $\phi$ in~\cite{Michele:2008qm} and
relating $N_{part} \simeq dN_{ch}/d\eta$ at $\eta$=2.9 as measured to
within 10\% by the $Si$ pixel telescope. Recent theoretical results on
thermal radiation from three major groups working in the field are
included for
comparison~\cite{vanHees:2006ng,Ruppert:2007cr,Renk:2008prc,RH:2008lp,Dusling:2007rh},
calculated absolutely (not normalized to the data). Results from a
fourth one~\cite{Bratkovskaya:2008bf} only cover the 2$\pi$ region and
are not yet available in $p_{T}$-differential form. The general
agreement between data and model results both as to spectral shapes
and to absolute yields is most remarkable, supporting the term
``thermal'' used throughout this paper.

At very low $p_{T}$, a strong rise towards low masses is seen in the
data, reflecting the Boltzmann factor, i.e. the Plank-like radiation
associated with a very broad, nearly flat spectral function. Only the
Hees/Rapp scenario~\cite{RH:2008lp} is able to describe this part
quantitatively, due to their particularly large contribution from
baryonic interactions to the low-mass tail of the $\rho$ spectral
function (as in ~\cite{Bratkovskaya:2008bf}). This was already
mentioned in connection with Fig.~\ref{fig3}, but is much more clearly
visible at low $p_{T}$ than without any $p_{T}$ selection. At higher
$p_{T}$, the influence of radial flow increasingly changes the
spectral shapes, and at very high $p_{T}$ all spectra appear
$\rho$-like. However, sizable differences between the different
theoretical scenarios also exist in this region. For example,
Hees/Rapp~\cite{RH:2008lp} use a hard-scattering $\rho$ which
contributes to fill up the $\rho$ region beyond the Cooper-Frye
freeze-out $\rho$. They also use an extrapolation of the Drell-Yan
process down to the photon point ($M$$\rightarrow$0). While this
contribution is small for the whole LMR region without $p_{T}$
selection, the low-mass/high-$p_{T}$ part is in their case completely
dominated by DY. The size of radial flow, the major issue here, also
varies between the three groups. It is recognizably too low in the
hydrodynamics scenario~\cite{Dusling:2007rh}, and maximal for the
fireball scenario of~\cite{Ruppert:2007cr,Renk:2008prc} tailored to
the NA60 hadron data.

\begin{figure*}[ht]
\begin{center}
\includegraphics*[width=0.45\textwidth]{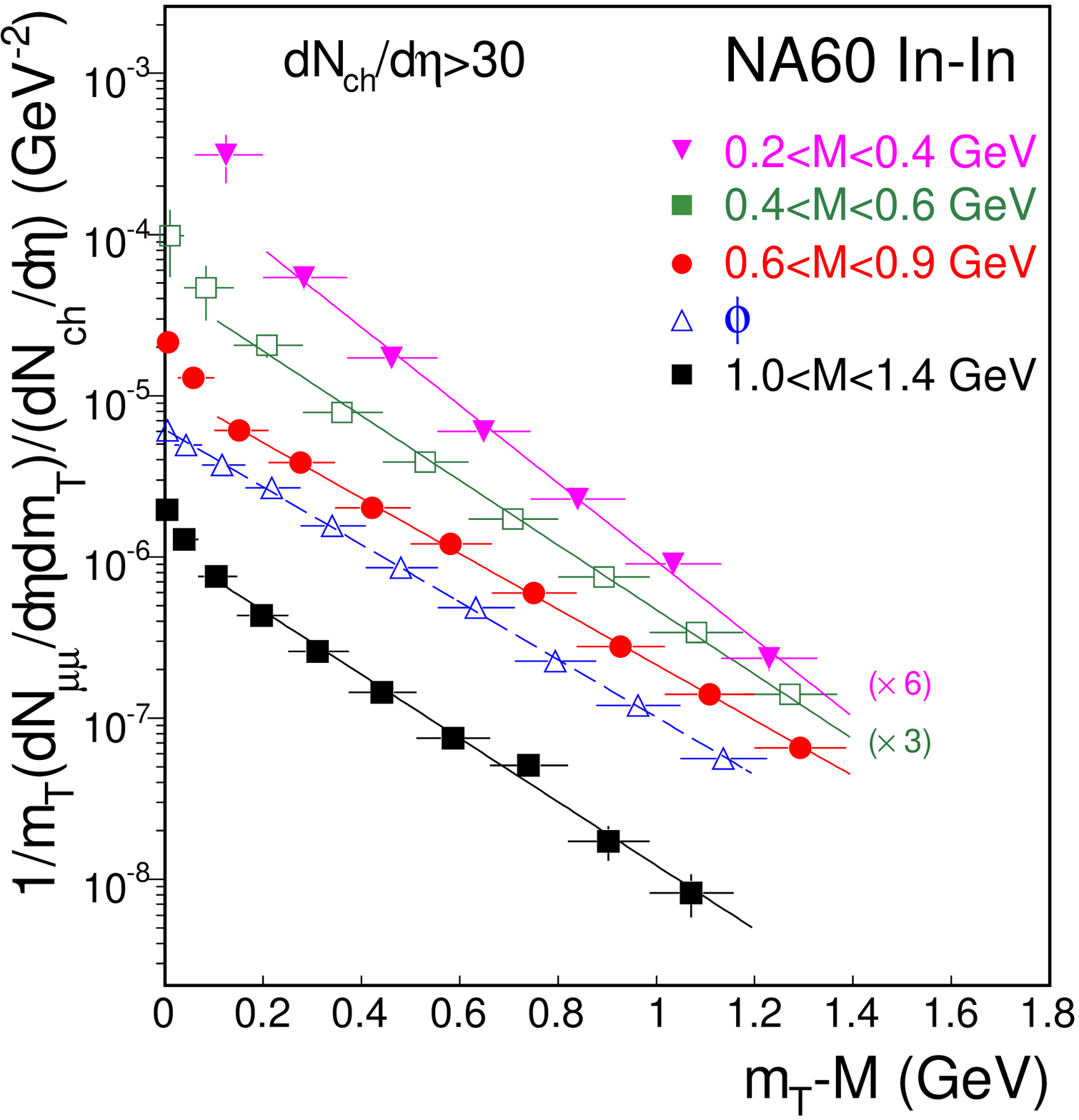}
\hspace*{1.0cm}
\includegraphics*[width=0.45\textwidth]{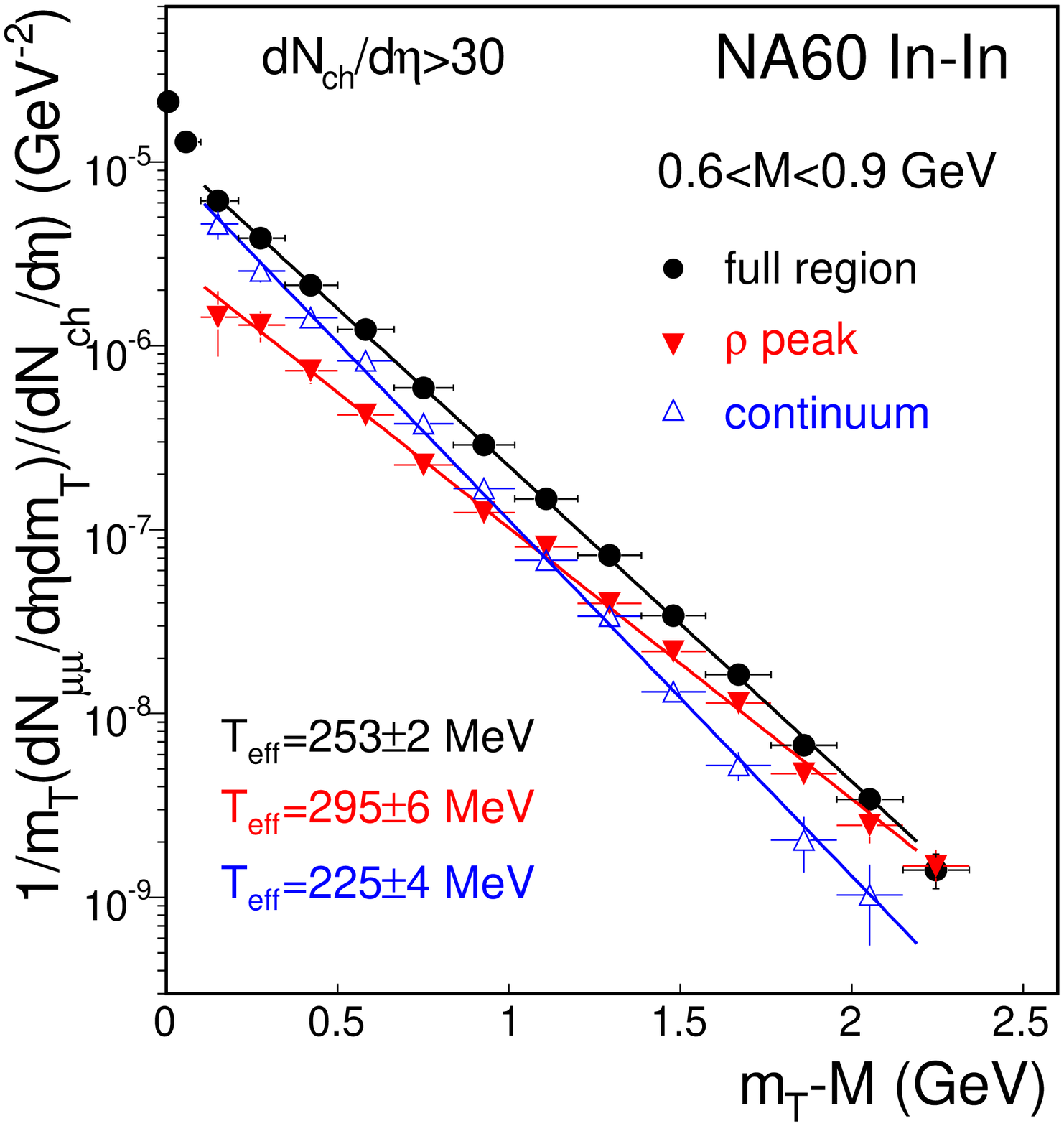}
\caption{Acceptance-corrected transverse mass spectra of the excess
dimuons for 4 mass windows and the $\phi$~\cite{Arnaldi:2007ru}
(left), and a decomposition into peak and continuum for the
$\rho$-like window (right, see text). Open charm is subtracted
throughout. The normalization in absolute terms is independent of
rapidity over the region measured. For error discussion
see~\cite{Arnaldi:2007ru}.}
   \label{fig5}
\end{center}
\end{figure*}

Fig.~\ref{fig5} (left) shows the centrality-integrated $m_{T}$
spectra, where $m_{T}$$=$$(p_{T}^{2} + M^{2})^{1/2}$, for four mass
windows; the $\phi$ is included for comparison. The ordinate is
absolutely normalized to $dN_{ch}/d\eta$ as in Fig.~\ref{fig4}. Apart
from a peculiar rise at low $m_{T}$ ($<$0.2 GeV) for the excess
spectra (not the $\phi$) which only disappears for very peripheral
collisions~\cite{Specht:2007ez,Arnaldi:2007ru}, all spectra are pure
exponentials. The rise is outside of any systematic errors as discussed
in~\cite{Arnaldi:2007ru}. The relative yield associated with it is
about 10-20\%, roughly independent of mass, which excludes a
connection to the low $p_{T}$ rise seen in pion $p_{T}$ spectra. The
absolute yield steeply decreases with mass, reminiscent of Dalitz
decays. However, a consistent physical interpretation is still
open. The lines in the exponential region are fits to the data with
the function $1/m_{T}dN/dm_{T}$ $\propto$
$exp(-m_{T}/T_\mathrm{eff})$, where the effective temperature
parameter $T_\mathrm{eff}$ is the inverse slope of the
distributions. For the excess data, the fits are restricted to the
range 0.4$<$$p_{T}$$<$1.8 GeV (roughly 0.1$<$$m_{T}-M$$<$1.2 GeV) to
exclude the increased rise at low m$_{T}$. Obviously, the slopes
depend on mass. Fig.~\ref{fig5} (right) shows a more detailed view
into the $\rho$-like mass window, exploiting the same side-window
method as used in connection with Fig.~\ref{fig9} below, to determine
the $p_{T}$ spectra separately for the $\rho$ peak and the underlying
continuum. All spectra are purely exponential up to the cut-off at
$p_{T}$=3 GeV, without any signs of an upward bend characteristic for
the onset of hard processes. Their slopes are, however, quite
different (see below).

\begin{figure}[b!]
\begin{center}
\includegraphics*[width=0.43\textwidth]{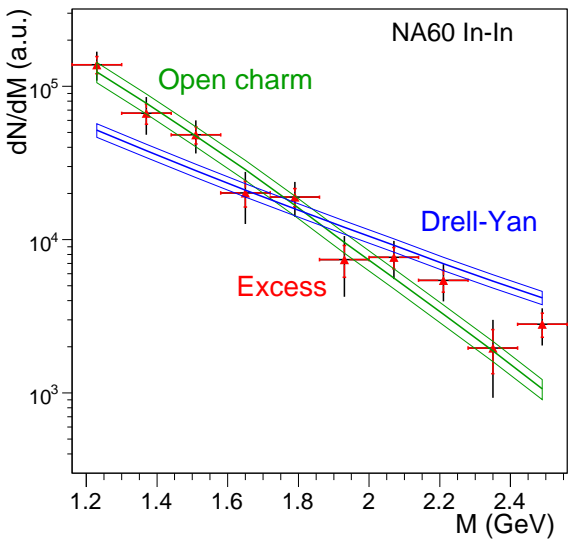}
\caption{Acceptance-corrected mass spectra of all three contributions
to the IMR spectrum: Drell-Yan, open charm and the excess
(triangles). The data are integrated over centrality.}
   \label{fig6}
\end{center}
\end{figure}

The central NA60 results in the IMR region~\cite{NA60:2008IMR} are
shown in Fig.~\ref{fig6}. The use of the $Si$-vertex tracker allows to
measure the offset between the muon tracks and the main interaction
vertex and thereby to disentangle prompt and offset dimuons from $D$
decays. The offset distribution is found to be perfectly consistent
with no charm enhancement, expressed by a fraction of $1.16\pm0.16$ of
the level expected from upscaling the NA50 results on the IMR in
$p$-$A$ collisions~\cite{NA60:2008IMR}. The observed excess is really
prompt, with an enhancement over Drell-Yan by a factor of
2.4$\pm$0.08. The excess can now be isolated in the same way as was
done in the LMR region, subtracting the measured known sources, here
DY and open charm, from the total data. Fig.~\ref{fig6} shows the
decomposition of the total into DY, open charm and the prompt
excess. The mass spectrum of the excess is quite similar to the shape
of open charm and much steeper than DY; this explains of course why
NA50 could describe the excess as enhanced open
charm~\cite{Abreu:2002rm}.

\begin{figure}[b!]
\begin{center}
\includegraphics*[width=0.43\textwidth]{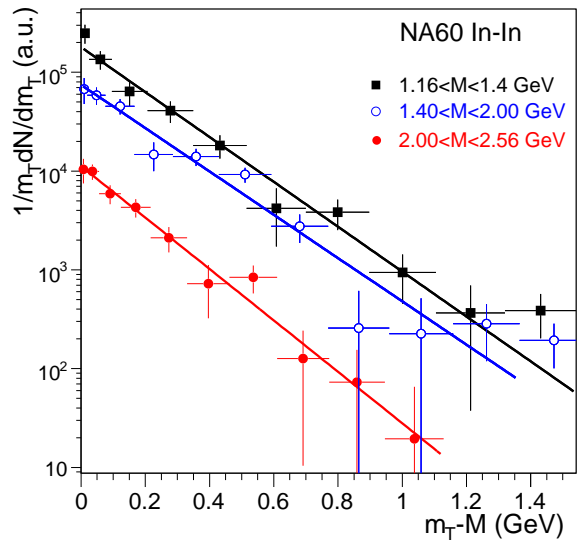}
\caption{Acceptance-corrected transverse mass spectra of the excess
dimuons for three mass windows. The data are integrated over
centrality. For error discussion see~\cite{NA60:2008IMR}.}
   \label{fig7}
\end{center}
\end{figure}

The transverse momentum spectra are also much steep\-er than
DY. Moreover, the spectra depend on mass and do not show the
factorization between mass and $p_{T}$ characteristic for DY, where a
common Gaussian distribution with a fixed sigma $k_{T}$=0.8 GeV
describes all $p_{T}$ spectra independent of mass. The transverse mass
spectra are shown in Fig.~\ref{fig7} for three consecutive mass
windows. All spectra are essentially exponential. However, the
steepening observed at very low $m_{T}$ in the lowest mass window,
seen already before for all masses in Fig.~\ref{fig5} including this
window, seems to be switched-off in the upper two mass windows. As in
Fig.~\ref{fig5}, the lines are exponential fits to the data,
restricted again to $p_{T}$$\geq$0.5 GeV to exclude the rise at low
$m_{T}$. The extracted inverse slope parameters are
199$\pm$21(stat)$\pm$3(syst), 193$\pm$16$\pm$2 and 171$\pm$21$\pm$3
MeV, respectively, i.e. about the same within the (rather large)
errors.

\begin{figure*}[ht]
\begin{center}
\includegraphics*[width=0.43\textwidth]{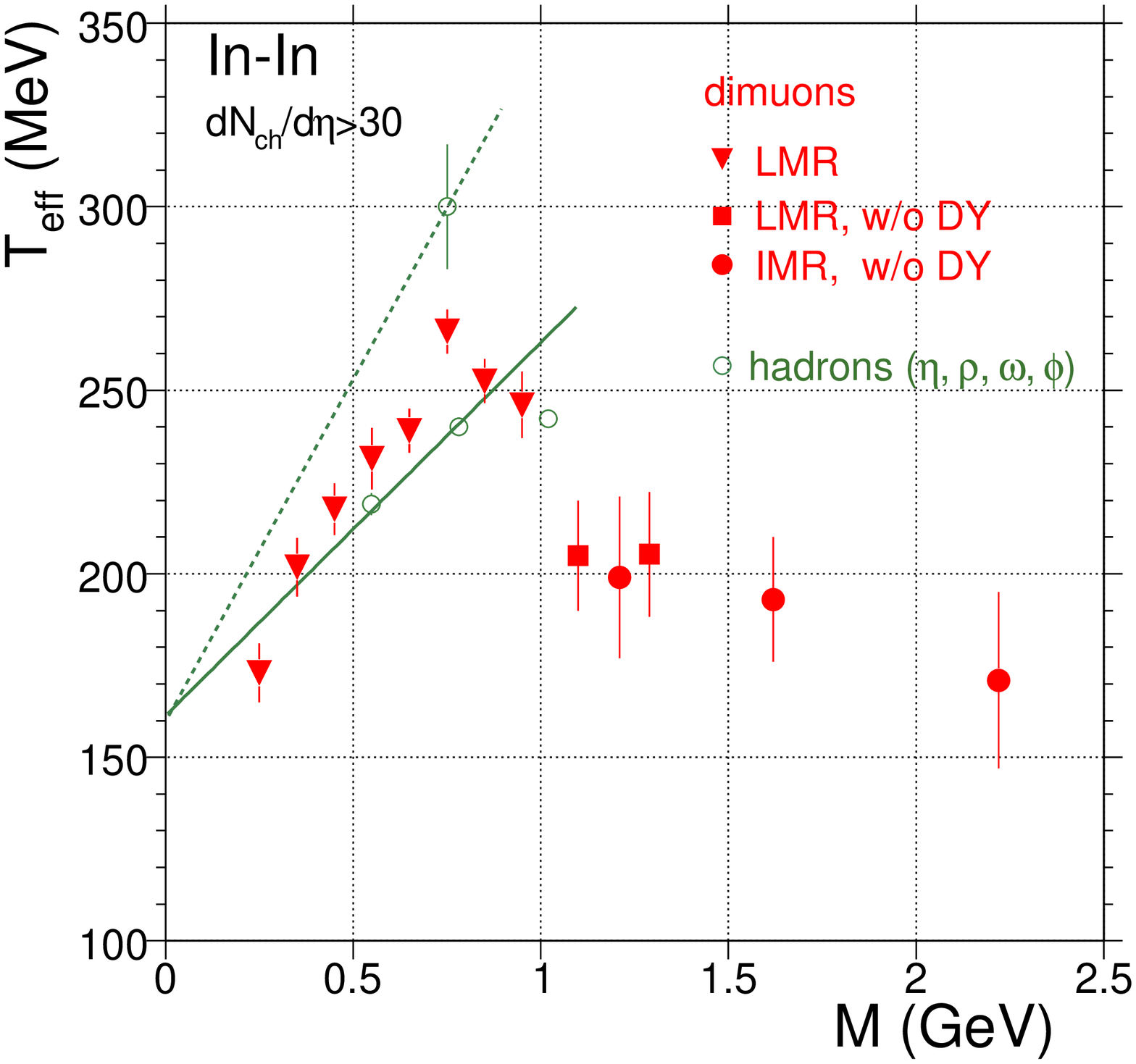}
\hspace*{1.0cm}
\includegraphics*[width=0.43\textwidth]{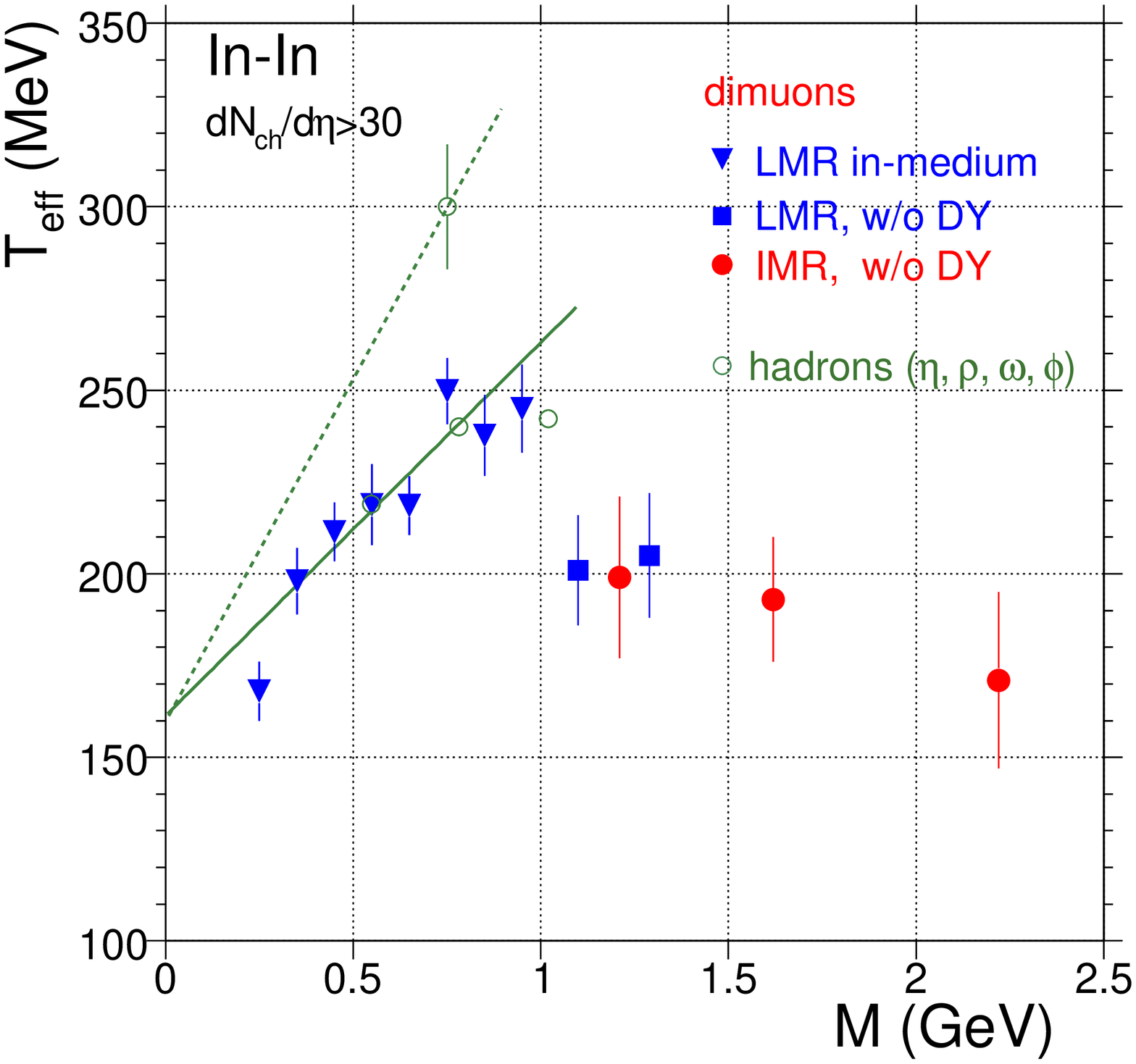}
\caption{Left: Inverse slope parameter $T_\mathrm{eff}$ vs. dimuon
mass for the combined LMR/IMR regions of the excess in comparison to
hadrons~\cite{Arnaldi:2007ru}. Right: Inverse slope parameters
$T_\mathrm{eff}$ for the ``pure'' in-medium part, obtained by
subtraction of the $\rho$-peak contribution from the total before the
fits are done (see text). Open charm is subtracted throughout. Errors
in the LMR part are purely statistical; the systematic errors are
mostly smaller than the statistical ones~\cite{Arnaldi:2007ru}. Errors
in the IMR part are total errors~\cite{NA60:2008IMR}.}
   \label{fig8}
\end{center}
\vglue 0.2cm
\end{figure*}

The inverse slope parameters $T_\mathrm{eff}$ extracted from the
exponential fits to the $m_{T}$ spectra are plotted in the left panel
of Fig.~\ref{fig8} vs. dimuon mass, unifying the data from the LMR and
IMR regions. In the LMR part, a finer binning is used than in
Fig.~\ref{fig5}, and the coarser-binned data (contained in
~\cite{Arnaldi:2007ru}) are left out for clarity. For M$<$1 GeV, a
correction for Drell-Yan pairs is not done, due to their small
contribution~\cite{Arnaldi:2007ru} and the intrinsic uncertainties at
low masses~\cite{RH:2008lp}. In the extended LMR analysis up to 1.4
GeV, the 2 (square) points are corrected, as are all points of the IMR
analysis (see above). In the region of overlap, the data are not
statistically independent. The hadron data for $\eta$, $\omega$ and
$\phi$ obtained as a by-product of the cocktail subtraction procedure
are also included in Fig.~\ref{fig8}, as is the single value for the
$\rho$-peak from the right panel of Fig.~\ref{fig5}. Interpreting the
latter as the freeze-out $\rho$ without in-medium effects, consistent
with all present theoretical
modeling~\cite{vanHees:2006ng,Ruppert:2007cr,Renk:2008prc,RH:2008lp,Dusling:2007rh},
all four hadron values together with preliminary $\pi^{-}$ data from
NA60 can be subjected to a simple blast wave
analysis~\cite{Damjanovic:2008ta}. This results in a reasonable set of
freeze-out parameters of the fireball evolution and suggests the
following consistent interpretation for the hadron and dimuon data
together. Maximal radial flow is reached by the $\rho$, due to its
maximal coupling to pions, while all other hadrons follow some
hierarchy in earlier freeze-out. The $T_\mathrm{eff}$ values of the
dimuon excess rise nearly linearly with mass up to the pole position
of the $\rho$, but stay always well below the $\rho$ line, completely
{\it consistent} with the expectations for {\it radial flow} of an
{\it in-medium hadron-like} source (here
$\pi^{+}\pi^{-}\rightarrow\rho$) decaying continuously into lepton
pairs.

This picture can still be refined. By modeling a $\rho$ with the
proper spectral shape and the $m_{T}$ spectrum as measured, its
contribution can be subtracted from the total measured distribution in
the full $M$-$m_{T}$ plane with the same side-window method as used
for Fig.~\ref{fig9} below and described
in~\cite{Damjanovic:2006bd}. The difference, essentially a continuum,
is then refit, resulting in $T_\mathrm{eff}$ values for the ``pure''
{\it in-medium} (continuum) part in the $2\pi$ region. The modified
plot is shown in the right panel of Fig.~\ref{fig8}. The appearance is
striking: all values are lower, but mostly so in the bin associated
with the $\rho$ pole (by $\sim$20 MeV). This makes the shape even more
sawtooth-like than before, and within errors the rise continues now up
to about 1 GeV.

Beyond the $2\pi$ region, the $T_\mathrm{eff}$ values of the excess
dimuons show a sudden decline by about 50 MeV down to the IMR values.
This decline is even more abrupt in the right panel of Fig.~\ref{fig8}
than in the left and obviously connected to the in-medium emission
itself, not to any peculiarities associated with the $\rho$
peak. Extrapolating the lower-mass trend set by a hadron-like source
to beyond 1~GeV, such a fast transition is extremely hard to reconcile
with emission sources which continue to be of dominantly hadronic
origin in this region. A much more natural explanation would be a
transition to a dominantly early, i.e. {\it partonic} emission source
with processes like $q\bar{q} \rightarrow \mu^{+}\mu^{-}$, for which
flow has not yet built up~\cite{Ruppert:2007cr,Renk:2008prc}. In this
sense, the present analysis may well represent the first data-based
evidence for thermal radiation of partonic origin in nuclear
collisions, overcoming parton-hadron duality in the yield description
on the basis of $M$-$p_{T}$ correlations as discussed in section 3.

Theoretically, the extension of the unified LMR and IMR results over
the complete $M$-$p_{T}$ plane places severe constraints on the
dynamical trajectories of the fireball evolution. Indeed all present
scenarios~\cite{vanHees:2006ng,Ruppert:2007cr,Renk:2008prc,RH:2008lp,Dusling:2007rh}
do not any longer rely on parton-hadron duality in the rates as
in~\cite{Rapp:1999zw}, but explicitly differentiate between hadronic
(mostly 4$\pi$) and partonic contributions in the IMR as already
discussed in connection with Fig.~\ref{fig3}. The partonic fraction
presently ranges from 0.65 for~\cite{RH:2008lp} (option EoS-B$^{+}$ as
used in Fig.~\ref{fig4}) to ``dominant''
in~\cite{Ruppert:2007cr,Renk:2008prc,Dusling:2007rh}. The exponential
shape of the experimental $m_{T}$ spectra is reproduced by the models,
consistent with the expectations for thermal radiation. However, due
to remaining uncertainties in the equation-of-state, in the fireball
evolution and in the role of hard processes~\cite{RH:2008lp}, a
quantitative description of the much more sensitive $m_{T}$-derivative
$T_\mathrm{eff}$ vs. $M$ in Fig.~\ref{fig8} is only slowly
emerging. In particular, the more recent results from the
authors of~\cite{Ruppert:2007cr,Renk:2008prc,Dusling:2007rh}, while very
encouraging in the description of the downward jump, are still
preliminary and have not yet been formally published in their final
form. A systematic comparison of several model results to the data in
Fig.~\ref{fig8} is therefore presently not possible.
\newpage

\begin{figure}[t!]
\centering
\includegraphics*[width=0.43\textwidth]{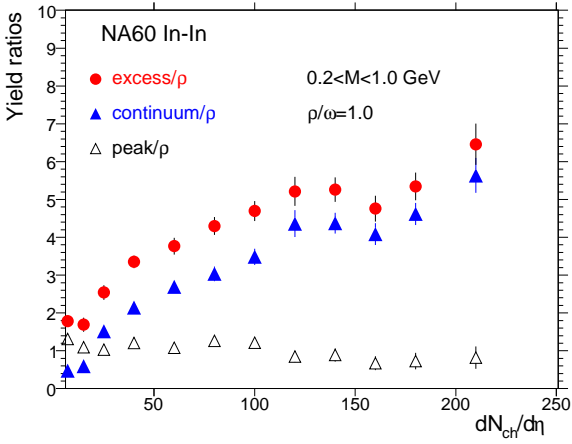}
\caption{Excess yield ratios for peak, continuum and total
vs. centrality for the mass window 0.2$<$$M$$<$1 GeV. Open charm is
subtracted throughout. No acceptance correction applied.}
   \label{fig9}
\end{figure}

\section{Centrality dependencies}
\label{sec:3}

All data presented so far correspond either to (nearly) minimum bias, 
selecting $dN_{ch}/d\eta$$>$30, or to the semicentral window,
selecting 110$<$$dN_{ch}/d\eta$$<$170; the results for the two
conditions are very close. However, an enormous amount of information
exists on the centrality dependence of practically every variable
discussed in this paper. We select two topics of particular relevance
here. 

The first one concerns the evolution of the shape of the excess mass
spectra, following Figs.~\ref{fig2}
and~\ref{fig3}. In~\cite{Damjanovic:2006bd}, we have used both an
$rms$ analysis and a more sensitive side-window method to determine
separately the yields of the peak and the underlying continuum. The
centrality dependence from the latter is shown in Fig.~\ref{fig9}:
peak, continuum and total excess yield in the mass interval
0.2$<$$M$$<$1.0 GeV, all normalized to the (fictitious) cocktail
$\rho$ with the assumption $\rho$/$\omega$=1 (like in $p$$p$). The
$\omega$ itself is directly measured, and its yield is found to be
proportional to $dN_{ch}/d\eta$. The continuum and the total show a
very strong increase, starting already in the peripheral region, while
the peak slowly decreases from $>$1 to $<$1. Recalling that
Fig.~\ref{fig9} is based on the excess mass spectra before acceptance
correction like in~Fig.~\ref{fig3}, roughly representing the full
$\rho$ spectral function, the excess/$\rho$ ratio can directly be
interpreted as the number of $\rho$ generations created by formation
and decay during the fireball evolution, including freeze-out: the
``$\rho$ clock'', frequently discussed in the past. It reaches up to
about 6 generations for central In-In collisions; selecting low
$p_{T}$ this number doubles.

\begin{figure}[t!]
\centering
\includegraphics*[width=0.43\textwidth]{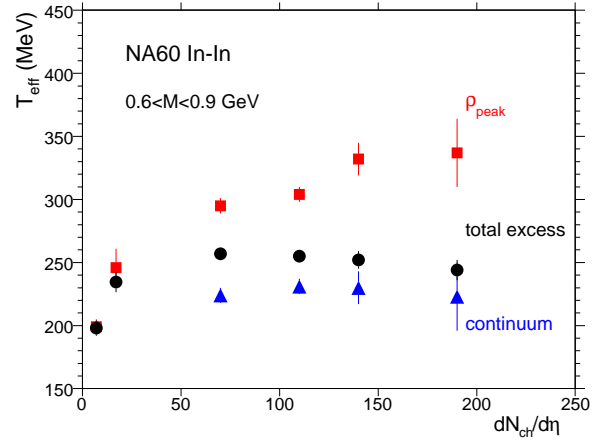}
\caption{Inverse slope parameter $T_\mathrm{eff}$ vs. centrality for
  continuum, peak and total in the mass window 0.6$<$$M$$<$0.9 GeV
  (see also~\ref{fig4}, right). Open charm is subtracted throughout.
  }
   \label{fig10}
\end{figure}

The second topic concerns the centrality dependence of the slope
parameter $T_\mathrm{eff}$ for the excess data in the $\rho$-like
window 0.6$<$$M$$<$0.9 GeV, following Fig.~\ref{fig4} (right). Based
as before on the side-window method~\cite{Damjanovic:2006bd}, the
results are shown in Fig.~\ref{fig10}, separately for the $\rho$ peak,
the continuum and the total excess. The peak is seen to show a very
strong rise, with hardly any saturation. However, the errors in the
more central data become quite large, reflecting the continuously
decreasing peak/total ratio as visible in Fig.~\ref{fig9}. Conversely,
continuum and total yield saturate much earlier.

The large gap in $T_\mathrm{eff}$ between the peak and the continuum,
seen already in Fig.~\ref{fig7}, has a much-debated interesting
physics origin. The $p_T$ spectrum of a thermal in-medium source is
softer by a Lorentz factor $M/E$$=$$1/\gamma$ compared to that of a
freely decaying freeze-out $\rho$~\cite{RH:2008lp}. On top, the
in-medium values of $T_\mathrm{eff}$ represent a temperature-flow
average, while the freeze-out $\rho$ receives the maximal flow. These
two effects contribute about equally to the total. The size of the gap
ultimately reaches 70-100~MeV, but closes towards peripheral
collisions. The $\omega$, with the same mass as the $\rho$, also
shows a large gap to the $\rho$ (compare Fig.~\ref{fig11}), which also
closes finally to zero for the lowest pp-like
window~\cite{Arnaldi:2007ru}.

\section{Evidence for $\omega$ in-medium modifications}
\label{sec:5}

While most of the historical discussion on light-flavor vector mesons
in hot and dense matter has concentrated on the short-lived $\rho$
($c\tau$~=~1.3~fm), the longer-lived $\omega$ (23~fm) and $\phi$
(46~fm) have received much less attention, since most of their
dilepton decays occur after thermal freeze-out. Within the NA50 LMR
analysis, the $\omega$ and $\phi$ have indeed consistently been
treated as "cocktail" particles and subtracted from the total (compare
Fig.~\ref{fig1}). However, in-medium effects are expected for the
(small) decay fraction inside the fireball, and these are actually
contained in the Hees/Rapp scenario (see~\cite{RH:2008lp} and earlier
references in there).

\begin{figure}[t!]
\begin{center}
\includegraphics*[width=0.43\textwidth]{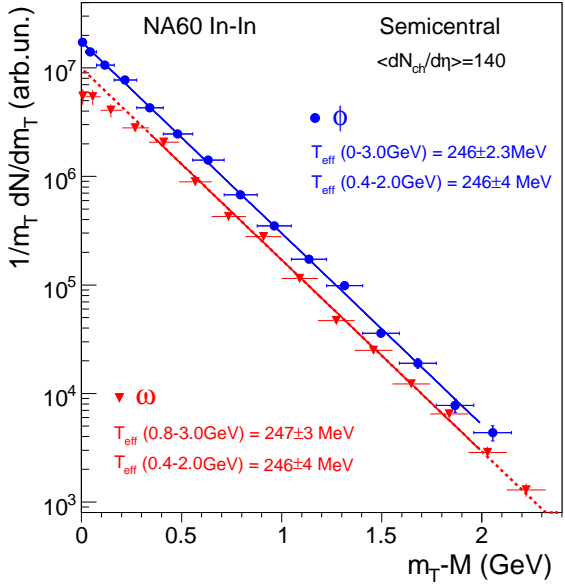}
\caption{Acceptance-corrected transverse mass spectra of the $\omega$
  and the $\phi$ for the semicentral bin. The $T_\mathrm{eff}$ values
  correspond to different fit ranges for the two particles. A
  depletion of the $\omega$ yield at low $m_T$ relative to the fit
  line can clearly be recognized.
}
   \label{fig11}
\end{center}
\end{figure}
%

NA60 has addressed the $\omega$ in a way directly coupled to the
cocktail subtraction procedure. Due to the high mass resolution, the
\textit{disappearance} of the yield at low $p_T$ out of the narrow
$\omega$ peak in the nominal pole position can sensitively be
detected. The \textit{appearance} of the yield elsewhere in the mass
spectrum, originating from a mass shift, or broadening or both, is
practically unmeasurable, due to the masking of the whole region by
the much stronger $\pi\pi \rightarrow \rho$ process, regenerating the
$\rho$. Sensitive experiments on $\omega$ in-medium effects with clear
clues as to their characteristics can therefore only be done in cold
nuclear matter experiments~\cite{Metag:2008xy}, where
$\rho/\omega~=~1$, but not in ultra-relativistic nuclear
collisions. The evidence for the disappearance of $\omega$'s in the
low $m_T$ region is shown in Fig.~\ref{fig11}.

As already mentioned in section 2, the $\omega$ and $\phi$ are
obtained as a byproduct of the cocktail subtraction procedure. The
data are fit with the usual $m_T$ exponential used before in
Figs.~\ref{fig4} and~\ref{fig6}. With respect to this reference line,
there is hardly any anomaly visible for the $\phi$, but quite some
loss for low-$p_T$ $\omega$'s. The loss can be quantified with respect
to the reference line, extrapolating down to zero. The fit parameters
$T_\mathrm{eff}$ for two different fit regions both for the $\omega$
and the $\phi$ show the definition of the reference line to be quite
uncritical. Forming the ratio data/reference line takes care of that
part of radial flow which does not seriously affect the exponential
slope.

\begin{figure}[t!]
\begin{center}
\includegraphics*[width=0.43\textwidth]{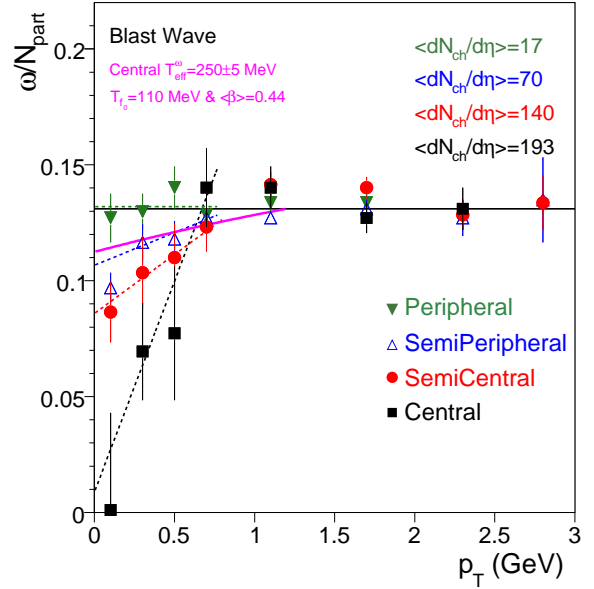}
\caption{ $p_T$ dependence of the $\omega$ yield with respect to the
fit line in Fig.~\ref{fig11}, absolutely normalized for the full phase
space yield, for different centralities. The solid line for
$p_T\le$~1~GeV shows the result from a blast-wave fit to the $\omega$
for central collisions. The dotted lines are only meant to guide the
eye. The errors are purely statistical. The systematic errors are
negligibly small compared to the statistical ones.}
   \label{fig12}
\end{center}
\end{figure}

The results are shown in Fig.~\ref{fig12}, absolutely normalized as
the full phase space ratio $\omega/N_{part}$. The effects of $\omega$
disappearance are quite striking: (i) a suppression of the relative
yield below the reference line only occurs for $p_T\le$~1~GeV; (ii)
there is a very strong centrality dependence of the suppression,
reaching down to $\le$~0.5 of the reference line (the errors become
huge for the central window, because the $\omega$ can then hardly be
recognized on top of the $\pi\pi$ processes at low $m_T$); (iii) the
suppression effects are much larger than expected for the spectral
distortions due to the blue shift from radial flow at low $m_T$; a
simulation in the basis of the blast wave parameters from
~\cite{Damjanovic:2008ta} shows at most 10$\%$ effects for central
collisions. Theoretical simulations addressing these results are not
yet available. It should be added that the same procedure applied to
the $\phi$ does describe the data solely on the basis of radial
flow. No effect beyond that can be recognized, within errors.

\section{Conclusions}
\label{conclude}

This paper, supplementing~\cite{Damjanovic:2008ta}, contains the most
comprehensive data set on excess dileptons above the known sources
which has so far become available through NA60. We have concentrated
here more than before on interpretational aspects, in particular on
the way, "parton-hadron duality" in the yields can be overcome by a
careful study of $M$-$p_T$ correlations. The data mediate a clear
conclusion on the dominance of partonic processes for $M$$>$1 GeV. A
systematic comparison with theoretical models reveals remaining
ambiguities in the modeling, but the overall agreement with the data
tends by now to support the same conclusion.

\end{document}